\documentclass[aps,prl,superscriptaddress,preprintnumbers,showpacs,a4paper,twoside,twocolumn,floatfix]{revtex4}

\usepackage{bm}
\usepackage{graphicx}
\usepackage{amsfonts}
\usepackage{amsmath}
\usepackage{amssymb}
\usepackage{amsthm}
\usepackage{bbm}
\usepackage{setspace}
\usepackage{helvet}

\newcommand{\kla}[1]{\left( #1 \right)}

\theoremstyle{plain}

\begin{document}

\title{Hawking Radiation from an Acoustic Black Hole on an Ion Ring}

\author{B. Horstmann}
\affiliation{Max-Planck-Institut f\"ur Quantenoptik, Hans-Kopfermann-Stra\ss e 1, 85748 Garching, Germany}
\author{B. Reznik}
\affiliation{Department of Physics and Astronomy, Tel Aviv University, Ramat Aviv 69978, Israel}
\author{S. Fagnocchi}
\affiliation{School of Physics and Astronomy, University of Nottingham, University Park, Nottingham NG7 2RD, United Kingdom}
\affiliation{SISSA, via Bonomea 265, 34136 Trieste, Italy}
\author{J. I. Cirac}
\affiliation{Max-Planck-Institut f\"ur Quantenoptik, Hans-Kopfermann-Stra\ss e 1, 85748 Garching, Germany}
\date{Garching, April 2009}


\begin{abstract}
In this Letter we propose to simulate acoustic black holes with ions in rings. If the ions are rotating with a stationary and inhomogeneous velocity profile, regions can appear where the ion velocity exceeds the group velocity of the phonons. In these regions phonons are trapped like light in black holes, even though we have a discrete field theory and a nonlinear dispersion relation. We study the appearance of Hawking radiation in this setup and propose a scheme to detect it.
\end{abstract}

\maketitle

In 1974 Hawking showed that the theory of quantum fields in curved spacetime predicts that, surprisingly, black holes emit thermal radiation \cite{Hawking74}. Unfortunately, the temperature of this radiation is too small to be detected for typical astrophysical black holes. Furthermore, the original theoretical derivation suffers from the problem that the wave equation is assumed to be valid on all scales, whereas the theory of quantum fields in curved space is assumed to break down at the Planck energy. Unruh showed that the Hawking effect is also manifested in analogous hydrodynamical systems which have a region of supersonic flow and hence a sonic horizon \cite{Unruh81}. Such analogous systems offer great advantages, since the effect can potentially be accessible to experiments. Moreover, its robustness can be examined based on the well known microphysics of the hydrodynamical systems. This will contribute to deepen our understanding of the Hawking effect also in gravitational black holes.

\begin{figure}[b]
\begin{center}
\includegraphics[width=77mm]{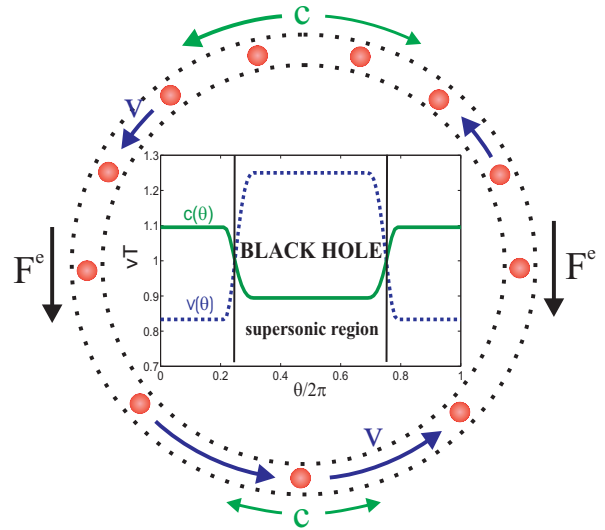}
\caption{Schematic depiction of the ion rotation with velocity $v\kla{\theta}$ and the phononic excitations with velocity $\pm c\kla{\theta}$, that depend on the varying ion spacing. The external force $F^e\kla{\theta}$ localized at the transition between the super- and subsonic region (de-)accelerates the ions. The inset shows a typical velocity profile with a black hole like region for $v\kla\theta>c\kla\theta$.}
\label{setup}
\end{center}
\end{figure}

The hydrodynamic analogy of gravitational spacetimes has inspired many proposals for experimental tests of Hawking radiation \emph{in continuous fields} in recent years \cite{Liberati05}, e.g., phonons in Bose-Einstein condensates \cite{Cirac00,Carusotto08,Lahav09}, Fermi gases \cite{Giovanazzi05}, superfluid Helium \cite{Jacobsen98}, slow light \cite{Leonhardt00,Unruh03}, and nonlinear electromagnetic waveguides \cite{Schutzhold05}. So far, no proposal has been physically implemented.

In the present work we show how to build an analog model of a black hole in an experimentally realizable system of ions. This is the first experimental proposal in which a discrete system is completely analyzed in the discrete limit (see also \cite{Schutzhold05}). A sublinear dispersion relation at high wave numbers naturally results from the discreteness of the physical system. This affects the trajectories of blue-shifted waves close to the event horizon \cite{Jacobsen91}. The dispersion relation is, additionally, nontrivial at low wave numbers because of the long range Coulomb force. We study how much this affects the appearing Hawking radiation. Explicit numerical calculations show that the Hawking effect is robust against such short scale modifications, e.g., for a continuous field with a sublinear dispersion relation \cite{Unruh99} and a discretized field on a falling lattice \cite{Jacobsen99}. Our proposal uses a parameter regime which is accessible in experiments at temperatures currently achieved. Thus, it could lead to the first experimental observation of Hawking radiation.

The main idea of our proposal can be summarized as follows. We are constructing a discrete analog of a hydrodynamical system with super- and subsonic regions in a quadrupole ring trap \cite{Drees64, Walther92} as schematically depicted in Fig. \ref{setup}. Ions rotate on a ring with cir\-cum\-ference $L$ with an inhomogeneous, but stationary velocity profile $v\kla\theta$. Thus the ions are inhomogeneously spaced. The necessary (de-)acceleration of the ions is guaranteed by additional electrodes exerting a force $F^e\kla \theta$ on the ions. The harmonic oscillations of the ions around this equilibrium motion are phonons with velocities $c\kla\theta\propto \kla{v\kla{\theta}}^{-1/2}$. Regions with large enough ion spacings and low enough velocities are supersonic; here phonons can only move into the direction of the ion flow and are trapped like light inside a black hole. We consider a system with a super- and a subsonic region. The border between these regions is analogous to a black hole horizon and will be shown to emit Hawking radiation.

In the following we explain the detailed setup of our proposal. The dynamics of $N$ ions with mass $m$ and charge $e$ are described by the Hamiltonian
\begin{equation}
 \mathcal{H}=-\sum_{i=1}^{N} \frac{4\pi^2\hbar^2}{2mL^2}\frac{\partial^2}{\partial\theta_i^2}+\sum_{i=1}^{N}V^e\kla{\theta_i}+V^c\kla{\theta_1,\dots,\theta_N}
\end{equation}
with the Coulomb potential $V^c$ and a local external potential $V^e\kla{\theta}$. Instead of specifying $V^e$ we will impose an angular velocity profile $v\kla{\theta}$ by fixing the equilibrium positions. The required $V^e$ is then determined through the difference between the ion acceleration $\ddot{\theta}_i^0\kla t$ and the Coulomb force $F_i^c\kla{\theta^0_1,\dots,\theta^0_N}$.

We impose the classical equilibrium positions
\begin{equation}
\theta_i^0(t)=g\kla{{i}/{N}+{t}/{T}},
\end{equation}
where $g$ maps the normalized indices $i/N\in [0,1]$ monotonically increasing onto the angles $\theta\in [0,2\pi]$ and is periodically continued.
The angular velocity profile is $v(\theta)=g'\kla{g^{-1}\kla{\theta}}/T$. We make the choice
\begin{equation}
\frac{g'(x)}{T}=\begin{cases}
       v_{min} & x\le \sigma-\gamma_1\\
	\beta+\alpha h\kla{\frac{x-\sigma}{\gamma_1}}	 & -\gamma_1<x-\sigma<\gamma_1\\
       v_{max} & \sigma+\gamma_1\le x\le 1-\sigma-\gamma_2\\
	\beta-\alpha h\kla{\frac{x-1+\sigma}{\gamma_2}} & -\gamma_2<x-\kla{1-\sigma}<\gamma_2\\
       v_{min} & x\ge 1-\sigma+\gamma_2
      \end{cases}
\end{equation}
with $\alpha=\frac{v_{max}-v_{min}}{2}$ and $\beta=\frac{v_{max}+v_{min}}{2}$ and $h(s)=15/8s-5/4s^3+3/8s^5$. $v\kla{\theta}=v_{max}$ is constant in the supersonic region $\sigma v_{min}T \lesssim \theta\lesssim 1-\sigma v_{min}T$ and $v\kla{\theta}=v_{min}$ constant outside of it ($v_{min}<2\pi/T<v_{max}$). We choose $v_{min}=2\pi\cdot 0.8\overline{3}/T$ and that the small transition regions $2\gamma_1,2\gamma_2$ contain $0.04N$ and $0.1N$ ions.


We treat small perturbations around the equilibrium motion $\theta_i\kla{t}=\theta_i^0\kla{t}+\delta\theta_i\kla{t}$
and expand the Hamiltonian to second order in $\delta\theta_i$
\begin{equation}
\label{Hamiltonian}
 \mathcal{H}=-\sum_{i=1}^{N} \frac{4\pi^2\hbar^2}{2mL^2}\frac{\partial^2}{{\partial\delta\theta}_i^2}+\frac{1}{2}\sum_{i\ne j}f_{ij}(t){\delta\theta}_i{\delta\theta}_j
\end{equation}
with the time dependent force matrix $f_{ij}$ and the canonical operators $\delta\theta_i$ and $-i\hbar\partial_{\delta\theta_i}$ describing the phononic oscillations of the ions. The quasifree quantum dynamics of this harmonic system are governed by the classical linear equations of motions for the first and second moments. We have checked that anharmonic effects do not significantly alter the results presented in this Letter. With Floquet analysis we find systems with stable and unstable dynamics in the interesting parameter regime similar to \cite{Cirac00}. In the presence of two horizons, black hole laser instabilities \cite{Corley99} may contribute to this. However, the instabilities become dominant at such late times that they are not important for the simulations presented here. Nevertheless, the examples given in this Letter represent stable systems.

In order to get some insight, we consider the limit of an infinite number of ions, where we can make an analogy with the standard Hawking effect for the scalar field $\Phi\kla{\theta^0_i\kla{t},t}=\delta\theta_i\kla{t}$ as observed in \cite{Unruh81}. Then for a slowly varying $v\kla{\theta}$ and Coulomb interactions between neighboring ions only the system Lagrangian becomes
\begin{equation}
 \mathcal{L}=\int d\theta\frac{\rho\kla{\theta}}{2}\left[\kla{\partial_t\Phi+v\kla\theta\partial_{\theta}\Phi}^2-\kla{iD\kla{\theta,-i\partial_\theta}\Phi}^2\right]
\end{equation}
with the dispersion relation $D\kla{\theta,k}=c\kla{\theta}k+\mathcal{O}\kla{k^3}$, the conformal factor $\rho\kla\theta=n\kla\theta \cdot mL^2/(2\pi)^2$, and the density $n\kla{\theta}=N/(v\kla\theta T)$. This scalar field satisfying a linear dispersion relation at low wave numbers with sound velocity $c\kla{\theta}=\sqrt{2(2\pi)^3 n\kla{\theta}e^2/(4\pi\epsilon_0\cdot mL^3)}$ is analogous to a massless scalar field in a black hole spacetime \cite{Unruh81}. Its quanta cannot escape a supersonic region with $v\kla{\theta}>c\kla{\theta}$ like photons trapped inside a black hole. The horizon of this analog model is located at $c\kla{\theta_H}=v\kla{\theta_H}$ with $\theta_H\approx \sigma v_{min}T$. Pairs of Hawking particles are emitted close to the black hole horizon with a black body distribution at the Hawking temperature
\begin{equation}
 \frac{k_B T_H}{\hbar}=\frac{1}{4\pi v}\frac{d}{d\theta}\kla{v^2-c^2}|_H=\frac{3}{4\pi T}\frac{g''\kla{g^{-1}\kla{\theta}}}{g'\kla{g^{-1}\kla{\theta}}}\bigr|_{\theta=\theta_H}.
\end{equation}
The first equality is derived in Ref. \cite{Unruh81}, the second one results from the explicit forms for $v\kla\theta$ and $c\kla\theta$.

Now, we can follow two different approaches to observe effects of Hawking radiation. In the first one we calculate the history of a final negative frequency pulse at small positive wave numbers leaving the black hole following the standard techniques of \cite{Unruh99,Jacobsen99}. It is found that the final low wave number pulse scatters off the horizon and originates from two pulses, which conserve the frequency. These pulses consisting of high absolute wave numbers have such a low group velocity that they are moving rightwards  in the lab frame, while being leftmoving (upstream) in the comoving frame. The positive or negative wave number pulse has negative or positive frequencies. A different situation described in \cite{Jacobsen99} appears, when the two low wavelength solutions of the frequency condition $\omega_0=vk\pm D\kla{k}$, with the typical final frequency $\omega_0$, lie outside the Brillouin zone. Then the incoming pulses are instead rightmoving (downstream) and the positive or negative frequency pulse is located at high positive or negative wave numbers. This effect is analogous to Bloch oscillations. The particle production associated with Hawking radiation is found through the comparison of the so-called Klein-Gordon norm $\mathcal{N}$ of the positive frequency initial pulse and the normalized final pulse. This initial pulse is found to reflect a thermal distribution at the Hawking temperature \cite{Unruh99, Jacobsen99}.

We now return to the discrete ion chain. Our numerical results for the propagation of a final pulse backwards in time in the discrete system of phonons on an ion ring are presented in the following. We first introduce the quantities necessary for this analysis. If the phononic excitations are localized in the flat subsonic region, the excitations $\delta\theta_i(t)$ and their canonic conjugate $\delta\dot\theta_i(t)$ can be expressed as modes $\delta\theta_k(t)$ and $\delta\dot\theta_k(t)$ with wave number $k$. Because of the finite system size only discrete wave numbers appear and thus the dispersion relation is also discrete. The positive and negative frequency part of these excitations are defined by
$2\delta\theta_k^\pm(t)=\delta\theta_k(t)\pm i \delta\dot\theta_k(t)/\omega_k$ and 
 $2\delta\dot\theta_k^\pm(t)=\delta\dot\theta_k(t) \mp i \omega_k \delta\theta_k(t)$.
The analysis of the particle production requires us to use the Klein-Gordon norm for these modes $\mathcal{N}_k^\pm=\delta\dot\theta_k^{\pm *}\delta\theta^\pm_k-\delta\theta_k^{\pm *}\delta\dot\theta^\pm_k$.
We start from the final pulses
\begin{equation}
\label{final}
 \delta\theta^s_k\kla{0}=k \cdot e^{-\kla{\frac{k-2\pi s}{40\pi}}^2}, \hspace{0.3 cm} s=1,\dots,20,
\end{equation}
to test different frequency ranges. We calculate its dynamics with the linear equations of motions following from the Hamiltonian \eqref{Hamiltonian}. Our proposal does not suffer from the problem reported in \cite{Unruh03} despite the steep dispersion at low wave numbers for long range Coulomb interactions. In our case, frequency conservation in the lab frame allows three pulses on the leftmoving branch of the dispersion relation exactly as in \cite{Unruh99,Jacobsen99}.

\begin{figure}[tbc]
\begin{center}
\includegraphics[width=80mm,height=60mm,angle=0]{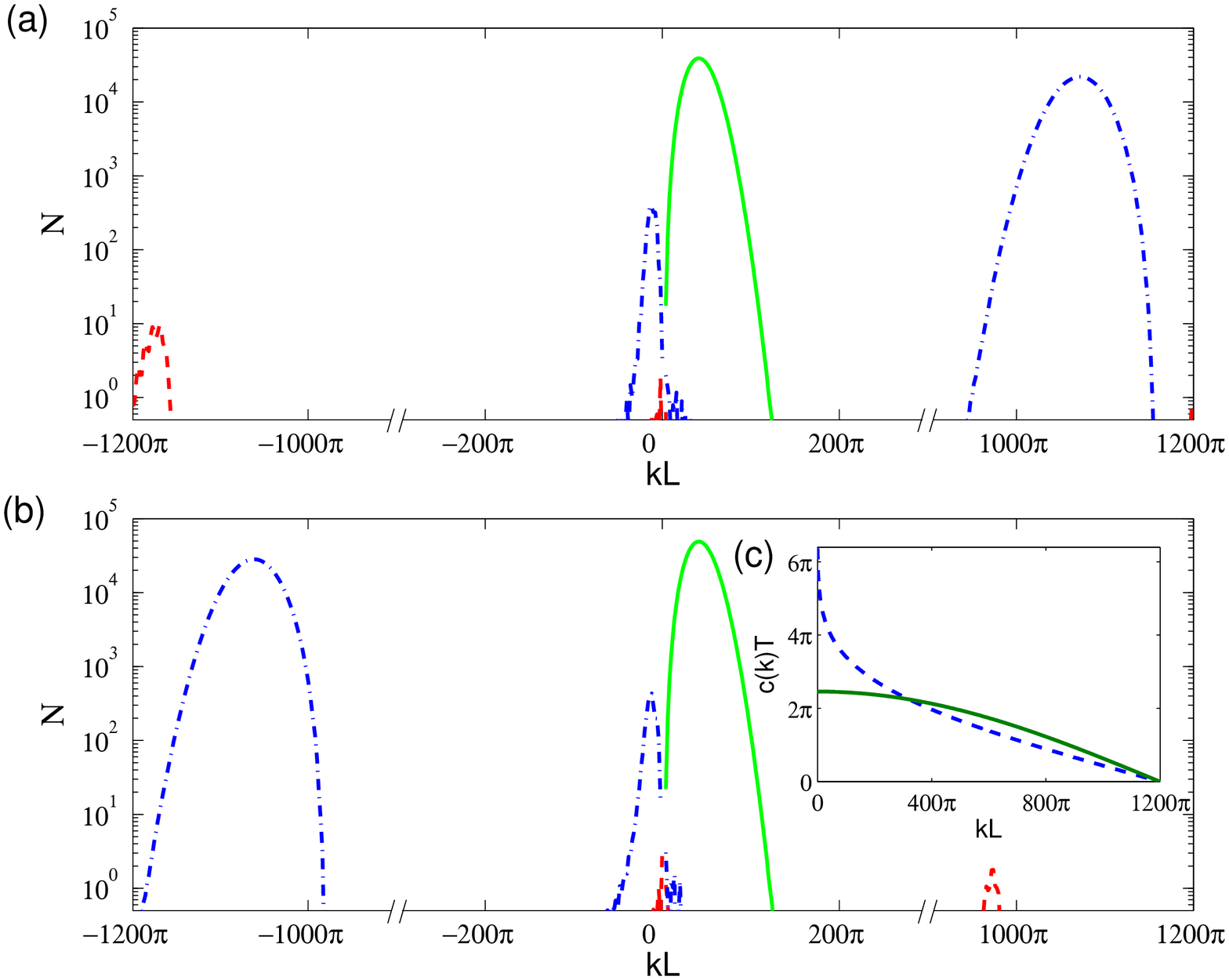}
\caption{Values of $\delta\theta_k(t)$ for propagation backwards in time starting from the final wave function in Eq. \eqref{final} with $s=5$. We use $\sigma v_{min}T=2\pi\times 0.375$, and $N=1000$. The final negative frequency pulses are depicted in green (solid line), the initial negative frequency pulses in blue (dash-dotted line), and the initial positive frequency pulses in red (dashed line). We depict the discrete norm distributions and dispersion relation as a continuous curve here. (a) Final ($t=0$) and initial ($t=-0.67T$) norm distribution of $\delta\theta_k(t)$; $e^2/4\pi\epsilon_0=1.2591/(2N)\cdot mL^3T^{-2}$. (b) Final ($t=0$) and initial ($t=-0.51T$) norm distribution of $\delta\theta_k(t)$ showing Bloch oscillation; $e^2/4\pi\epsilon_0=2.0004/(2N)\cdot mL^3T^{-2}$. (c) Phononic group velocity in the flat subsonic region as a function of $k$ for full interactions (blue dashed line) and nearest-neighbor interactions only (green solid line); $e^2$ as in (a).}
\label{fourierspace}
\end{center}
\end{figure}

Our results agree mainly with those of Refs. \cite{Unruh99,Jacobsen99} as shown in Fig. \ref{fourierspace}. We find the two previously mentioned high wave number pulses. In addition, we observe a weak rightmoving pulse with small negative wave numbers (see Fig. \ref{fourierspace}), whose presence agrees with approximate frequency conservation in the lab frame. This pulse could be a consequence of the finite initial and final excitation probability of ions outside the flat subsonic region. For small ion velocities the effect reported in \cite{Jacobsen99} analogous to Bloch oscillations is observed. Then the main incoming pulses are rightmoving.

We further compare the Klein-Gordon norm of the positive frequency initial pulse with the prediction for thermal radiation \cite{Unruh99,Jacobsen99}. If only nearest-neighbor Coulomb interactions are considered, the relative difference between these norms is lower than $\epsilon=0.01$ for $N=1000$ ions. For the long range Coulomb interactions, a conservative estimate yields $\epsilon\le 0.2$, which is in good agreement with thermal radiation.  This conclusion is also confirmed by the analogous spectrally dissolved comparison (see \cite{Unruh99}). Here the definition of a global Hawking temperature $T_H$ is difficult because of the nonlinear dispersion relation at low wave numbers. We calculate $T_H$ from the discrete dispersion relation at low wave numbers. The value of $\epsilon$ can be explained with the inequality between the group and the phase velocity. Therefore, we have shown that Hawking radiation qualitatively persists even in a fundamentally discrete system with long range interactions and a logarithmically diverging group velocity at low wave numbers [see Fig. \ref{fourierspace}c].

\begin{figure}[t]
\begin{center}
\includegraphics[width=86mm,angle=0]{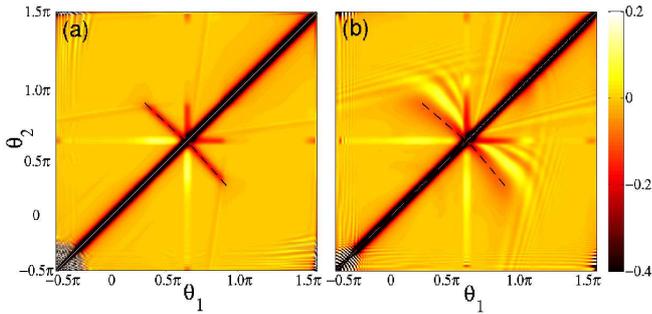}
\caption{Correlations $C_{ij}(t)$ at time $t$ with real space lab frame positions. We consider $N=1000$ ions, $\sigma v_{min}T=2\pi\times 0.25$, and $\tau=0.05T$. (a) Only nearest-neighbor interactions are considered with $e^2/4\pi\epsilon_0={1.127}/{(2N)}\cdot{mL^3}{T^{-2}}$ and $t=0.5T$. (b) Full Coulomb interactions are considered with $e^2/4\pi\epsilon_0={0.2453}/{(2N)}\cdot{mL^3}{T^{-2}}$ and $t=0.6T$.}
\label{densitydensity}
\end{center}
\end{figure}

For an experimental proof for Hawking radiation on the ion ring, we propose to measure correlations between the different regions on the ion ring after the creation of a black hole following \cite{Carusotto08}. Starting from a thermal state of the excitations around homogeneously spaced ions at rest, the ions are brought into rotation and in a subsequent small time interval $\tau$ a supersonic region is created. This is done by reducing the subsonic fluid velocity $v_{min}$ in a Gaussian way, while leaving the average rotation velocity constant. We are analyzing the normalized angle-angle correlations
\begin{equation}
  C_{ij}={\langle p_i p_j\rangle}\cdot T/(\hbar m).
\end{equation}
In Fig. \ref{densitydensity} we show the results at a fixed time starting from the ground state. Correlations between the super- and subsonic regions are created close to the black hole horizon and are moving away from it as expected. These correlations correspond to the pair creation of Hawking particles. We interpret their pure existence as a signature for Hawking radiation. Figure \ref{densitydensity}a shows the simulation for interactions between neighboring ions only \cite{excitations}, Fig. \ref{densitydensity}b for long range Coulomb interactions. The correlations behave similarly in both cases. We can tune and predict with good accuracy the angle and the propagation velocity of the correlation features based on the group velocities, i.e., the dispersion relation, and the ion velocities. The dashed lines in Fig. \ref{densitydensity} show these predictions for the direction of the main correlation signal.

We now discuss the parameter regime necessary in an experiment. Because the ion velocity must lie in the same order of magnitude as the phonon velocity in the proposed experimental setup, $e^2/4\pi\epsilon_0\approx 1/\kla{2N} mL^3/T^2$ is required. This is satisfied for $N=1000$ singly charged $^9Be$ ions with an average spacing of $L/N=10\mu m$ and a rotation frequency of the ions of $\omega_{ion}=2\pi\times 5.5kHz$. The Hawking temperature in this system is $k_BT_H/\hbar\approx 5/T\approx 2\pi\times 4.4kHz$. We have checked that the correlations of Fig. \ref{densitydensity} remain present with initial temperatures 2 orders of magnitude above the Hawking temperature \cite{Carusotto08}. At too high initial temperatures the emerging correlations are classical. The quantum nature of the correlations at low enough temperatures can be studied through the entanglement in the form of the logarithmic negativity \cite{Vidal02} which is created between the inside and the outside of the black hole.  We detect increasing entanglement in time between two regions adjacent to the horizon for initial temperatures up to 10 times the Hawking temperature. These parameters and this temperature should be reached in current experiments. We have checked that it is possible to control the equilibrium motion of the ions and to measure the correlations in the oscillations around them with sufficient accuracy. Note that it has been demonstrated long ago how to trap ions in quadrupole ring traps \cite{Drees64} and measure their arrangement \cite{Walther92}. Thus, the proposed experiment will allow for the measurement of signatures of Hawking radiation for acoustic black holes.

To conclude we have shown that it is possible to realize an acoustic black hole in a system of ions rotating on a ring. In this system thermally distributed Hawking radiation is expected, and we proposed a realistic experiment to measure signatures of Hawking radiation. We expect ring traps to be extremely useful for future quantum simulations, especially of translationally invariant systems.

We would like to thank D. Porras for many fruitful discussions and the Nanosystems Initiative Munich and the German-Israeli Science foundation for financial support. B.R. acknowledges the Israel Science Foundation grant 920/09. S.F.'s research is supported by Anne McLaren fellowship.


\begin{thebibliography}{99}
\bibitem{excitations} Excitations created at the white hole horizon have been reduced by a wider transition region at this horizon. In an experiment the magnitude and velocity of these excitations can be watched through careful measurements.
\end{thebibliography}
\end{document}